\documentclass[aps, prl, groupedaddress, amsmath, amssymb, reprint]{revtex4-1}

\pdfoutput=1

\usepackage{graphicx}
\usepackage{hyperref}

\begin{document}

\title{Comment on ``Scalings for radiation from plasma bubbles" \cite{PoP2010Thomas}}

\author{S. Corde}
\author{A. Stordeur}
\author{V. Malka}
\affiliation{Laboratoire d'Optique Appliqu\'ee, ENSTA ParisTech - CNRS UMR7639 - \'Ecole Polytechnique, Chemin de la Huni\`ere, 91761 Palaiseau, France}

\begin{abstract}
Thomas has recently derived scaling laws for X-ray radiation from electrons accelerated in plasma bubbles, as well as a threshold for the self-injection of background electrons into the bubble \cite{PoP2010Thomas}. To obtain this threshold, the equations of motion for a test electron are studied within the frame of the bubble model, where the bubble is described by prescribed electromagnetic fields and has a perfectly spherical shape. The author affirms that \textit{any} elliptical trajectory of the form $x^{\prime}{^2}/\gamma_p^2+y^{\prime}{^2}=R^2$ is solution of the equations of motion (in the bubble frame), within the approximation $p_y^\prime{^2}/p_x^\prime{^2}\ll1$. In addition, he highlights that his result is different from the work of Kostyukov \textit{et al.} \cite{PRL2009Kostyukov}, and explains the error committed by Kostyukov-Nerush-Pukhov-Seredov (KNPS).

In this comment, we show that numerically integrated trajectories, based on the same equations than the analytical work of Thomas, lead to a completely different result for the self-injection threshold, the result published by KNPS \cite{PRL2009Kostyukov}. We explain why the analytical analysis of Thomas fails and we provide a discussion based on numerical simulations which show exactly where the difference arises. We also show that the arguments of Thomas concerning the error of KNPS do not hold, and that their analysis is mathematically correct. Finally, we emphasize that if the KNPS threshold is found not to be verified in PIC (Particle In Cell) simulations or experiments, it is due to a deficiency of the model itself, and not to an error in the mathematical derivation.
\end{abstract}

\maketitle

Authors of Ref. \cite{PoP2010Thomas} and Ref. \cite{PRL2009Kostyukov} have
considered a model in which the bubble is described by prescribed
electromagnetic fields and has a perfectly spherical shape in the laboratory
frame, whose radius is $r_b$ and velocity is $v_p=c\sqrt{1-1/\gamma_p^2}$. They
obtained different thresholds for electron self-injection into the bubble.
Whereas Thomas argues that an error has been committed in the work of KNPS,
leading to wrong conclusions, we will show in this comment that the conclusions
of KNPS are correct (in the frame of the considered model) and that the
mathematical derivation of Thomas is erroneous. We begin by demonstrating that
there is no elliptical solution for the equations of motion, whatever the
initial conditions. Then, we explain why the arguments of Thomas concerning the error of KNPS do not hold, and we present numerical results showing agreement with the KNPS threshold. Finally, we provide a discussion based on numerical simulations which show exactly why considering the trajectory as elliptical leads to erroneous conclusions. We give qualitative arguments which highlight that the considered model could be too simple to quantitatively describe the self-injection physics.

In the following, we use the prime to indicate quantities defined in the bubble
rest frame, as opposed to quantities defined in the laboratory frame. In
addition, quantities are normalized by the choice $m_e=c=e=\omega_p=1$ where
$\omega_p$ is the plasma frequency. Derivatives with respect to
the electron proper time $\tau$ are indicated with a dot: $\dot{A}=dA/d\tau$.

\section{Elliptical trajectory}
In our conventions, the system of equations given by Eqs. (12) and (15) of
Ref. \cite{PoP2010Thomas} (equations of motion in the bubble frame) is written
\begin{eqnarray}
 \label{eq1}
 \ddot{x}^\prime &=& -\frac{1}{2\gamma_p}(\gamma^\prime x^\prime - \gamma_p^2\dot{y}^\prime y^\prime), \\
 \label{eq2}
 \ddot{y}^\prime &=& -\frac{\gamma_p}{2}(\gamma^\prime + \dot{x}^\prime) y^\prime.
\end{eqnarray}
From these equations, Eq. (18) of Ref. \cite{PoP2010Thomas} can be established (with a minus sign instead of a plus sign in the l.h.s, and a factor $m_e$ in the r.h.s) and is written
\begin{eqnarray}
 \label{eq3}
 \ddot{x}^\prime - \ddot{y}^\prime \frac{x^\prime}{\gamma_p^2y^\prime} &=& \frac{1}{4\gamma_p}\frac{d}{d\tau}(x^\prime{^2}+\gamma_p^2y^\prime{^2}).
\end{eqnarray}
Note that, while Thomas made use of the  approximation $p_y^\prime{^2}/p_x^\prime{^2}\ll1$ to derive Eq. (\ref{eq3}), this last equation can be derived without this approximation, such that, according to him, elliptical trajectories are not only approximate solutions (in the sense $p_y^\prime{^2}/p_x^\prime{^2}\ll1$) but exact solutions to the equations of motion.

If Eqs. (\ref{eq1}) and (\ref{eq2}) imply Eq. (\ref{eq3}), the reverse is false. Providing initial conditions ($x^\prime(0), y^\prime(0), \dot{x}^\prime(0), \dot{y}^\prime(0)$) are known, there are an infinite number of solutions for Eq. (\ref{eq3}), while only one for the system (\ref{eq1})+(\ref{eq2}). Thomas states that \textit{``This equation is satisfied by \underline{any} trajectory of the form $x^{\prime}{^2}/\gamma_p^2+y^{\prime}{^2}=R^2$''}. There is an infinite number of elliptical trajectories of this type, and 
they can be parametrized by $x^{\prime}(\tau)=\gamma_pR\cos\theta$,
$y^{\prime}(\tau)=R\sin\theta$ where $\theta$ is a function of $\tau$
(specifying a particular solution). According to the Thomas' affirmation,
\textit{any} elliptical trajectory is solution of Eq. (\ref{eq3}), which means in
mathematical terms: $\forall\theta\in\mathcal{C}^2,
(x^{\prime}(\tau)=\gamma_pR\cos\theta, y^{\prime}(\tau)=R\sin\theta)\in
S_{(3)}$, where $S_{(3)}$ is the solution space of Eq. (\ref{eq3}). Inserting this
parametrization into Eq. (\ref{eq3}) shows that terms in $\dot{\theta}^2$ and in
$\ddot{\theta}$ do not cancel out, and that \textit{any} trajectory of the form
$x^{\prime}{^2}/\gamma_p^2+y^{\prime}{^2}=R^2$ is not solution of Eq. (\ref{eq3}).
Instead, we obtain a differential equation for $\theta$, which has an unique
solution $\theta_s(\tau)$ providing that the initial conditions $(\theta(0),
\dot{\theta}(0))$ are known. We note $x_s^{\prime}=\gamma_pR\cos\theta_s$ and
$y_s^{\prime}=R\sin\theta_s$. By derivation, the trajectory
($x_s^{\prime},y_s^{\prime}$) is solution of Eq. (\ref{eq3}) and satisfies
$x^{\prime}{^2}/\gamma_p^2+y^{\prime}{^2}=R^2$. However, because Eq. (\ref{eq3}) is not
equivalent to Eqs. (\ref{eq1})+(\ref{eq2}), ($x_s^{\prime},y_s^{\prime}$) is \textit{a priori}
not a solution of the system (\ref{eq1})+(\ref{eq2}). To show that ($x_s^{\prime},y_s^{\prime}$)
is effectively not a solution of the system (\ref{eq1})+(\ref{eq2}), its
expression can be inserted in Eqs. (\ref{eq1}) and (\ref{eq2}). Here we propose a simpler
demonstration based on a Taylor expansion of the solution of Eqs. (\ref{eq1})+(\ref{eq2}) around
the initial time $\tau=0$:
\begin{eqnarray}
\nonumber
 x^\prime &=& x^\prime(0) + \dot{x}^\prime(0)\tau +
  \frac{\ddot{x}^\prime(0)}{2}\tau^2 + \frac{\dddot{x}^\prime(0)}{6}\tau^3 \\
 \label{eq4}
 &+& \frac{\ddddot{x}^\prime(0)}{24}\tau^4 + o(\tau^4) , \\
\nonumber
 y^\prime &=& y^\prime(0) + \dot{y}^\prime(0)\tau +
\frac{\ddot{y}^\prime(0)}{2}\tau^2 + \frac{\dddot{y}^\prime(0)}{6}\tau^3 \\
 \label{eq5}
 &+& \frac{\ddddot{y}^\prime(0)}{24}\tau^4 + o(\tau^4) .
\end{eqnarray}
An elliptical trajectory has to satisfy the following relation:
\begin{eqnarray}
 \label{eq6}
 x^\prime\dot{x}^\prime + \gamma_p^2y^\prime\dot{y}^\prime &=& 0,
\end{eqnarray}
at all time $\tau$. The initial conditions, compatible with Eq. (\ref{eq6}), are
$x^\prime(0)=0$, $y^\prime(0)=r_b$, $\dot{x}^\prime(0)=p_{x0}^\prime$,
$\dot{y}^\prime(0)=0$, where $p_{x0}^{\prime}$ is the only free parameter. To
derive the second, third and fourth derivative of $x^\prime$ and $y^\prime$ at
$\tau=0$ from Eqs. (\ref{eq1}) and (\ref{eq2}), the following relation is useful (it can be
obtained from Eqs. (8) and (12) of Ref. \cite{PoP2010Thomas}):
\begin{eqnarray}
\label{eq7}
 \frac{d}{d\tau}(\gamma^\prime+\dot{x}^\prime) &=& -\frac{1}{2\gamma_p} (\gamma^\prime+\dot{x}^\prime) x^\prime.
\end{eqnarray}
We obtain from Eqs. (\ref{eq1}) and (\ref{eq2}), and using Eq. (\ref{eq7}):
\begin{eqnarray}
\nonumber
 \ddot{x}^\prime(0) &=& 0, \\
\nonumber
 \dddot{x}^\prime(0) &=& -\frac{1}{2\gamma_p}\big[\gamma_0^\prime p_{x0}^\prime+\frac{\gamma_p^3r_b^2}{2}(\gamma_0^\prime+p_{x0}^\prime)\big], \\
\nonumber
 \ddddot{x}^\prime(0) &=& 0, \\
\nonumber
 \ddot{y}^\prime(0) &=& -\frac{\gamma_pr_b}{2}(\gamma_0^\prime+p_{x0}^\prime), \\
\nonumber
 \dddot{y}^\prime(0) &=& 0, \\
 \label{eq8}
 \ddddot{y}^\prime(0) &=& \frac{r_b}{4}(\gamma_0^\prime+p_{x0}^\prime)p_{x0}^\prime+\frac{\gamma_p^2r_b}{4}(\gamma_0^\prime+p_{x0}^\prime)^2, 
\end{eqnarray}
where $\gamma_0^\prime{^2} = 1 + p_{x0}^\prime{^2}$. We can insert this Taylor expansion of the solution of the equations of motion (\ref{eq1}) and (\ref{eq2}) in the relation for the elliptical trajectory, Eq. (\ref{eq6}), to check if the real solution is elliptical or not, in the limit $\tau\ll1$. Identifying each order of expansion gives:
\begin{eqnarray}
\nonumber
 &\textrm{Order 0}:& \qquad 0 = 0, \\
 \label{eq9}
 &\textrm{Order 1}:& \qquad p_{x0}^\prime{^2} - \frac{\gamma_p^3r_b^2}{2}(\gamma_0^\prime+p_{x0}^\prime) = 0 , \\
\nonumber
 &\textrm{Order 2}:& \qquad 0 = 0, \\
\nonumber
 &\textrm{Order 3}:& \qquad -\frac{\gamma_0^\prime p_{x0}^\prime{^2}}{3\gamma_p}
- \frac{3\gamma_p^2r_b^2}{24}(\gamma_0^\prime+p_{x0}^\prime)p_{x0}^\prime \\
\label{eq10}
 && \qquad +\frac{\gamma_p^4r_b^2}{6}(\gamma_0^\prime+p_{x0}^\prime)^2 = 0.
\end{eqnarray}
Eqs. (\ref{eq9}) and (\ref{eq10}) have to be satisfied simultaneously, whereas there is only one
free parameter $p_{x0}^\prime$. These equations are in fact incompatibles, they
can not be satisfied simultaneously. For example, if we consider the limit $|p_{x0}^\prime|\gg1$, we obtain $p_{x0}^\prime=-\gamma_p\sqrt[3]{r_b^2/4}$ from Eq. (\ref{eq9}), which is not solution of Eq. (\ref{eq10}).

\begin{figure}
   \centering
   \includegraphics[width=8.5cm]{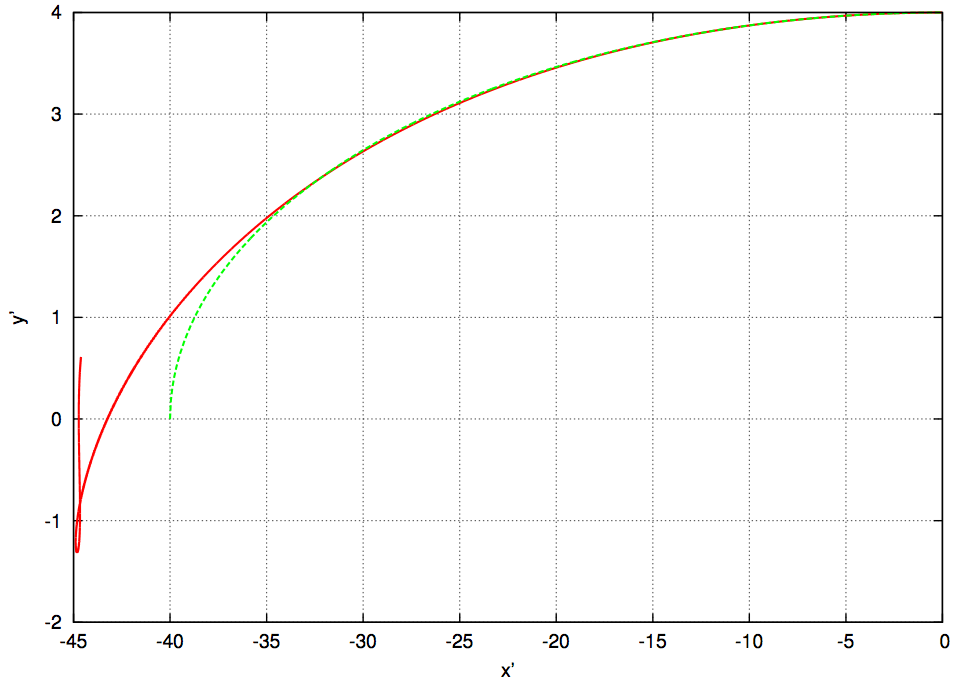}
   \caption{Electron trajectory for $\gamma_p=10$  and for the initial
conditions $x^\prime(0)=0$, $y^\prime(0)=4$,
$\dot{x}^\prime(0)=p_{x0}^\prime$, $\dot{y}^\prime(0)=0$ [where $p_{x0}^\prime$
is chosen as the solution of Eq. (\ref{eq9})]. The numerical solution of Eqs. (\ref{eq1}) and
(\ref{eq2}) is in solid red line, while the elliptical trajectory is in
dashed green line.}
   \label{fig1}
\end{figure}

\begin{figure}
   \centering
   \includegraphics[width=8.5cm]{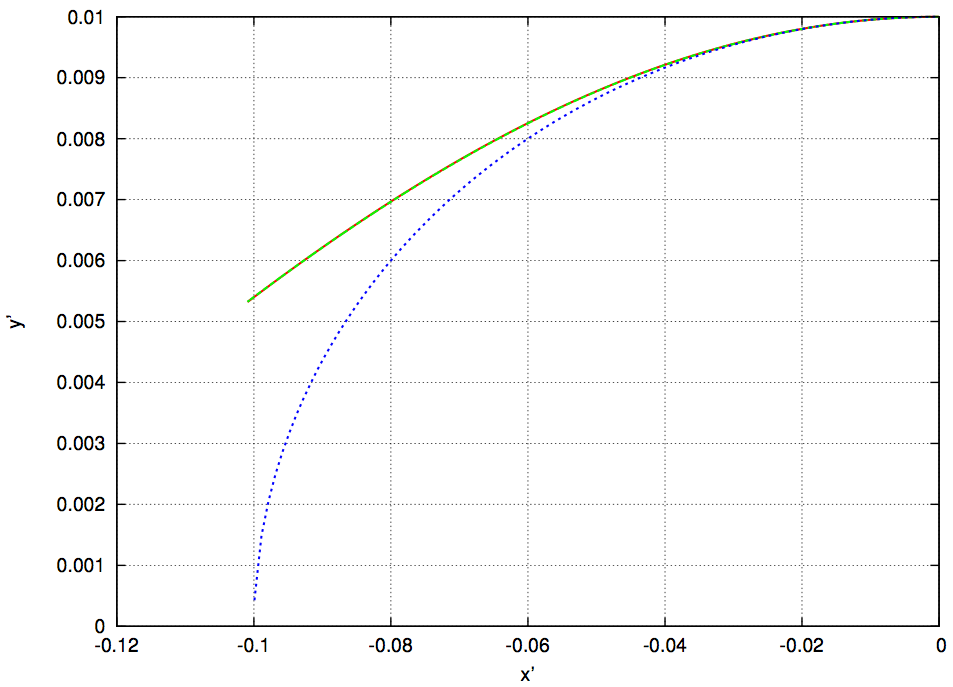}   
   \caption{Electron trajectory for $\gamma_p=10$  and for the initial
conditions $x^\prime(0)=0$, $y^\prime(0)=0.01$,
$\dot{x}^\prime(0)=p_{x0}^\prime$, $\dot{y}^\prime(0)=0$ [where $p_{x0}^\prime$
is chosen as the solution of Eq. (\ref{eq9})]. The numerical solution of Eqs. (\ref{eq1}) and
(\ref{eq2}) is in solid red line, while the Taylor expansion of the solution, given by
Eqs. (\ref{eq4}), (\ref{eq5}) and (\ref{eq8}), is in dashed green line (superposed to the red line). The
elliptical trajectory is in
dotted blue line. Trajectories are plotted up to $\tau=0.5$.}
   \label{fig2}
\end{figure}

In addition to this analytical analysis, a numerical integration of the
equations of motion can be performed to verify if the trajectory can be
elliptical, providing the correct choice of initial conditions. The value of
$p_{x0}^\prime$ is chosen as the solution of Eq. (\ref{eq9}), so that the trajectory is
effectively elliptical to the lowest order in $\tau$. A numerically
integrated trajectory is displayed on Fig. \ref{fig1} for the parameters $r_b=4$
and $\gamma_p=10$.
It is easily seen that the real trajectory does not follow an ellipse. We
checked that errors due to finite time step and numerical truncation were
negligible; varying the time step or the level of truncation has no effect on
the result. The trajectory is also found not to be sensitive to initial
conditions, for the time scale of interest.
Moreover, we performed a cross-verification of both the analytical
and numerical calculations. The Taylor expansion is valid only for $\tau\ll1$, and the time required for the electron to reach the back of the bubble is, in orders of magnitude, $\tau\sim\gamma_pr_b/|p_{x0}|^\prime\sim r_b^{1/3}$. Therefore we need $r_b^{1/3}\ll1$ for the expansion to be valid on the length scale of interest (the bubble extension). On Fig. \ref{fig2} is represented both the
analytical Taylor expansion [given by Eqs. (\ref{eq4}), (\ref{eq5}) and (\ref{eq8})] and the numerically
integrated trajectory [solution of Eqs. (\ref{eq1}) and (\ref{eq2})], up to $\tau=0.5$, for
$r_b=0.01$ and $\gamma_p=10$. The choice $r_b=0.01$ is only used here to perform a verification between analytical and numerical calculations (but this case does not have any physical relevance since the bubble model makes sense only for $a_0>2$, i.e. for $r_b>2\sqrt{2}$).
Both trajectories are very close to each other, confirming both the analytical
and numerical calculations.

We conclude that the real trajectory is not elliptical, whatever the initial
conditions. We will see in the Discussion section why incorrectly considering
the trajectory as elliptical leads to erroneous conclusions.

\section{On the error of KNPS}
Thomas argues that in the work of KNPS, the approximations made in Eqs. (4) to
(7) of Ref. \cite{PRL2009Kostyukov} are too restrictive and fail to correctly
predict the self-injection threshold. This can be easily understood by regarding
at Eq. (6) of Ref. \cite{PRL2009Kostyukov}: $X$ necessarily decreases, even when
$P_x\rightarrow\infty$ (in the notation of Ref. \cite{PRL2009Kostyukov},
$X=\xi/r_b=(x-v_pt)/r_b$, $P_x=p_x/r_b^2$). Such equations can not describe the
injection, since when the electron is injected, $X$ is increasing. Nevertheless,
Eqs. (4) to (7) of Ref. \cite{PRL2009Kostyukov} are only used to obtain the
numerical coefficient $P_x\simeq1.1$ at the moment where $p_y=0$ for the first
time, and to insert it into Eq. (3). The KNPS threshold is based on the
conservation of the hamiltonian $\mathcal{H}$ between the initial time and the
critical time where $p_y=0$ for the first time, and no approximation is needed
in this approach. From that, Eq. (3) of Ref. \cite{PRL2009Kostyukov} is
established. Moreover, a simple analysis in orders of magnitude of the equations
of motion Eqs. (1) and (2) of Ref. \cite{PRL2009Kostyukov} shows that
$p_x\propto r_b^2$. Inserting this behavior in Eq. (3) of Ref.
\cite{PRL2009Kostyukov} demonstrates the
self-injection threshold of Eq. (9) in Ref. \cite{PRL2009Kostyukov}, but without
the numerical coefficient. The
numerical coefficient can then be evaluated by drastically simplifying the
equations of motion, as done by KNPS. In reality, the coefficient could have a
very weak dependance on the parameters $r_b$ and $\gamma_p$, since the real
equations depend on them.

Therefore, the arguments of Thomas concerning the error of KNPS do not hold,
and the semi-analytical derivation of KNPS is correct.

\section{Numerical threshold}
\begin{figure}
   \centering
   \includegraphics[width=8.5cm]{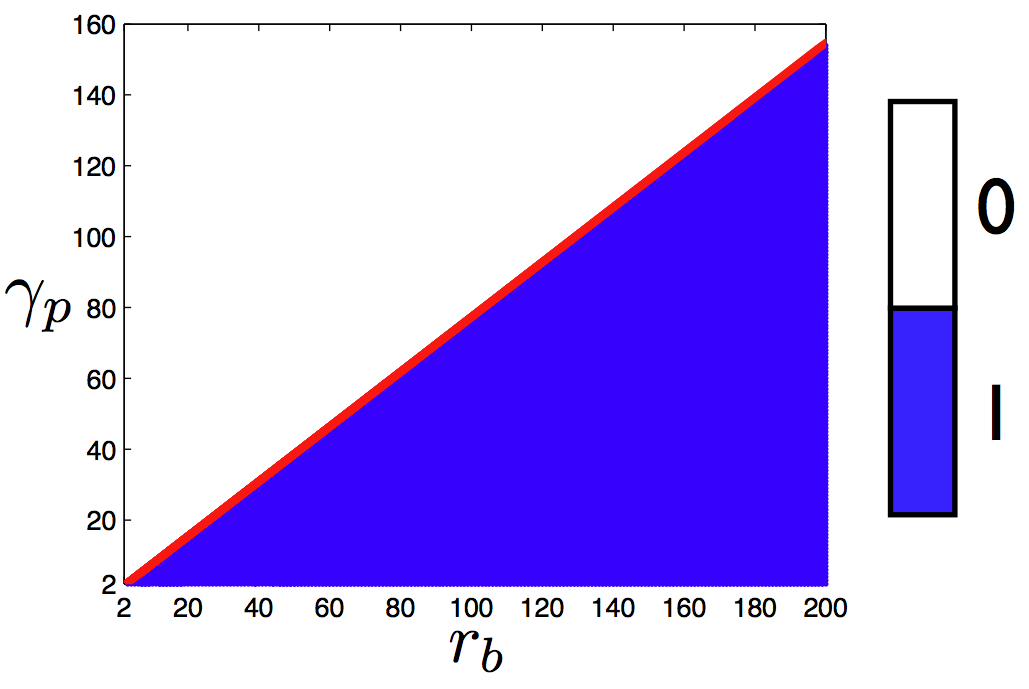}   
   \caption{For each value of $r_b$ and $\gamma_p$, a value of 1 is displayed if
the electron is injected and 0 otherwise. The initial conditions are
$x^\prime(0)=0$, $y^\prime(0)=r_b$, $\dot{x}^\prime(0)=-\gamma_pv_p$,
$\dot{y}^\prime(0)=0$ (electron at rest in the laboratory frame), and a
trajectory is considered injected if $r\leqslant r_b$ at all time steps.
The frontier between injected and non-injected trajectories follows a line of
equation $\gamma_p=0.77\;r_b$ (plotted in red line), whose numerical coefficient
is very close to the value obtained by KNPS \cite{PRL2009Kostyukov}.}
   \label{fig3}
\end{figure}

In order to verify the mathematical validity of the self-injection thresholds
proposed either by Thomas or KNPS, we have integrated the equations of motion
and scanned all the parameter space $(r_b,\gamma_p)$, with the same initial
conditions as Thomas or KNPS, i.e. $x^\prime(0)=0$, $y^\prime(0)=r_b$,
$\dot{x}^\prime(0)=-\gamma_pv_p$, $\dot{y}^\prime(0)=0$ (electron at rest in the
laboratory frame). The electron is considered to be injected if
$r^2=x^{\prime}{^2}/\gamma_p^2+y^{\prime}{^2} \leqslant r_b^{2}$ at all time
steps. Note that if the electron escapes the bubble before $y=0$, it will never
come back inside, so that imposing the condition $r\leqslant r_b$ only when
$y\leqslant0$ gives the same result. The numerical result is presented on Fig.
\ref{fig3} and is in agreement with the work of KNPS. In the frame of the model
considered by KNPS and Thomas (with initial conditions corresponding to an
electron at rest in the laboratory frame), the threshold is written
$r_b>1.30\gamma_p$.

\section{Discussion}
In this comment, we have analyzed the mathematical validity of the results
proposed either by Thomas and KNPS, considering a particular model where the
bubble is considered perfectly spherical and described by prescribed
electromagnetic fields. However, it is clear that such a simple model can
potentially fail to correctly describe the physical mechanisms present in the
blow-out regime of laser-plasma interaction. In the work of KNPS
\cite{PRL2009Kostyukov}, only one PIC simulation has been performed, while
several simulations with very different parameters should be performed to
confirm the linear self-injection threshold. In addition, we highlight that,
while Thomas considers the parameter $\gamma_p$ as a direct function of the
electron plasma density $\gamma_p\propto n_e^{-1/2}$, it should be instead
considered as the bubble back gamma factor which can be much lower due to the
time evolution of bubble. Indeed, Kostyukov and co-workers included the rate of
bubble expansion in the bubble back gamma factor \cite{NJP2010Kostyukov} and
found similar results that those of Kalmykov \textit{et al.}
\cite{PRL2009Kalmykov}, which have explicitly studied electron injection inside
a time-dependent bubble. This bubble back gamma factor has to be properly taken
into account if we want to verify the linear self-injection threshold of KNPS by
PIC simulations.

\begin{figure}
   \centering
   \includegraphics[width=8.5cm]{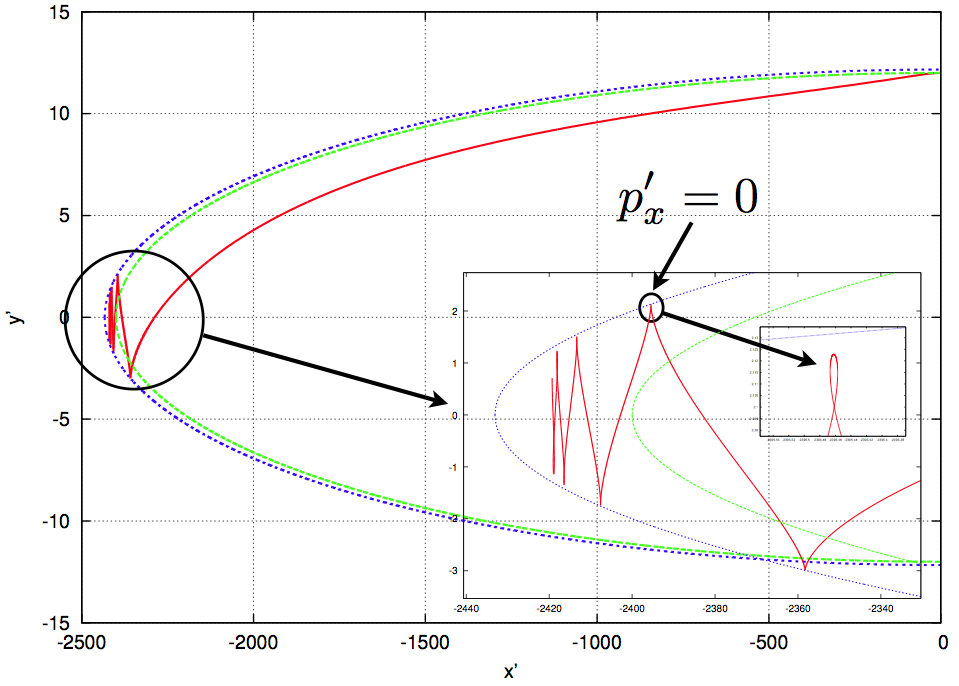}   
   \caption{Numerically integrated electron trajectory for $\gamma_p=200$  and
for the initial conditions $x^\prime(0)=0$, $y^\prime(0)=12$,
$\dot{x}^\prime(0)=-\gamma_pv_p$, $\dot{y}^\prime(0)=0$ (electron at rest in the
laboratory frame), in solid red line. The ellipse of equation $r=r_b$ is in
dashed green line, and the ellipse of equation $r=r_\textrm{max}$ is in
dotted blue line. The inset displays a zoom of the back of the bubble and shows
the point where $p_x^\prime=0$ for the first time. At that point,
$r\simeq r_\textrm{max}\geqslant r_b$ and $|x^{\prime}|/\gamma_p\leqslant r_b$.}
   \label{fig4}
\end{figure}

We have seen that, contrary to the Thomas' affirmation, there is no elliptical
solution for Eqs. (\ref{eq1}) and (\ref{eq2}). The difference between the result of Thomas and
the linear self-injection threshold of KNPS can be understood as follows. In the
bubble frame, the hamiltonian is written
$\mathcal{H}^\prime=\gamma^\prime-\phi^\prime$, where
$\phi^\prime\simeq2\gamma_p\phi=-\gamma_pr^2/4$, $\phi^\prime$ and $\phi$
being the scalar potential respectively in the bubble and laboratory frame.
The model is time-independant in the bubble frame, therefore
$\mathcal{H}^\prime$ is conserved
and $\gamma^\prime+\gamma_pr^2/4=\gamma_0^\prime+\gamma_pr_b^2/4$ [this is Eq.
(9) of Ref. \cite{PoP2010Thomas}]. Because $\gamma^\prime\geqslant1$, there is a
maximal value for $r$, which is given by $r_\textrm{max}\simeq
r_b(1+4/r_b^2)^{1/2}$ for $\gamma_p\gg1$ and $\gamma_0^\prime\simeq\gamma_p$.
For large values of $r_b$, $r_\textrm{max}$ becomes very close to $r_b$, and if a
small error is committed, an electron can be seen injected while it is not in
the frame of the considered model.
Thomas used Eq. (\ref{eq7}), considering $\gamma^\prime$ as a constant (which is
equivalent as saying that the trajectory is elliptical, according to the
conservation of $\mathcal{H}^\prime$), and studied the motion in terms of the
$x^\prime$ and $p_x^\prime$ variables. In reality, $\gamma^\prime$ is not
constant along the trajectory, which induces some degree of error in the
calculation of the relation between $x^\prime$ and $p_x^\prime$, given by Eq.
(20) of Ref. \cite{PoP2010Thomas}. In fact, the difference in the result
of Thomas arises when he considered an electron to be injected if $x^{\prime}
\geqslant -\gamma_pr_b$ when $p_x^{\prime}=0$, implicitly assuming an
elliptical trajectory, for which $y^\prime=0$ and $|x^{\prime}|/\gamma_p=r$ when
$p_x^\prime=0$, such that the condition $|x^{\prime}|/\gamma_p\leqslant r_b$ is
equivalent to $r\leqslant r_b$. But because the trajectory is not
elliptical, when $p_x^\prime=0$, $y^\prime\neq0$ and $|x^{\prime}|/\gamma_p\neq
r$ such that even if $r_b < r \leqslant r_\textrm{max}$ (the electron is
not injected), we can have $|x^{\prime}|/\gamma_p\leqslant r_b$. For large $r_b$,
even if $r= r_\textrm{max}$ when $p_x^\prime=0$, because $r_\textrm{max}$ is
very close to $r_b$, $|x^{\prime}|/\gamma_p$ will be smaller than $r_b$ due to
the non-zero value of $y^\prime$. For example, for $r_b=12$ and $\gamma_p=200$,
according to Thomas the electron is injected, while it is not according to
KNPS. Figure \ref{fig4} displays the corresponding trajectory (with
initial conditions for an electron at rest in the laboratory frame) and the
ellipses of equation $r=r_b$ and $r=r_\textrm{max}$.
At the moment where $p_x^\prime=0$ for the first time, $r\simeq
r_\textrm{max} > r_b$ (it is considered as non-injected by KNPS),
whereas $|x^{\prime}|/\gamma_p\leqslant r_b$ (it is injected according to the
Thomas' criterion). This example highlights that because $r_b$ and
$r_\textrm{max}$ are very close to each other, a small error in the
derivation or in the criterion can considerably change the conclusion (injected
or non-injected). In addition, in that case, it is clear that $\gamma^\prime$ is
not constant at all, since it almost attains $\gamma^\prime=1$
(when $r\simeq r_\textrm{max}$) and it attains very large values
$\gamma^\prime\gg\gamma_0^\prime$ during the period where the electron is inside
the bubble. 

\begin{figure}
   \centering
   \includegraphics[width=8.5cm]{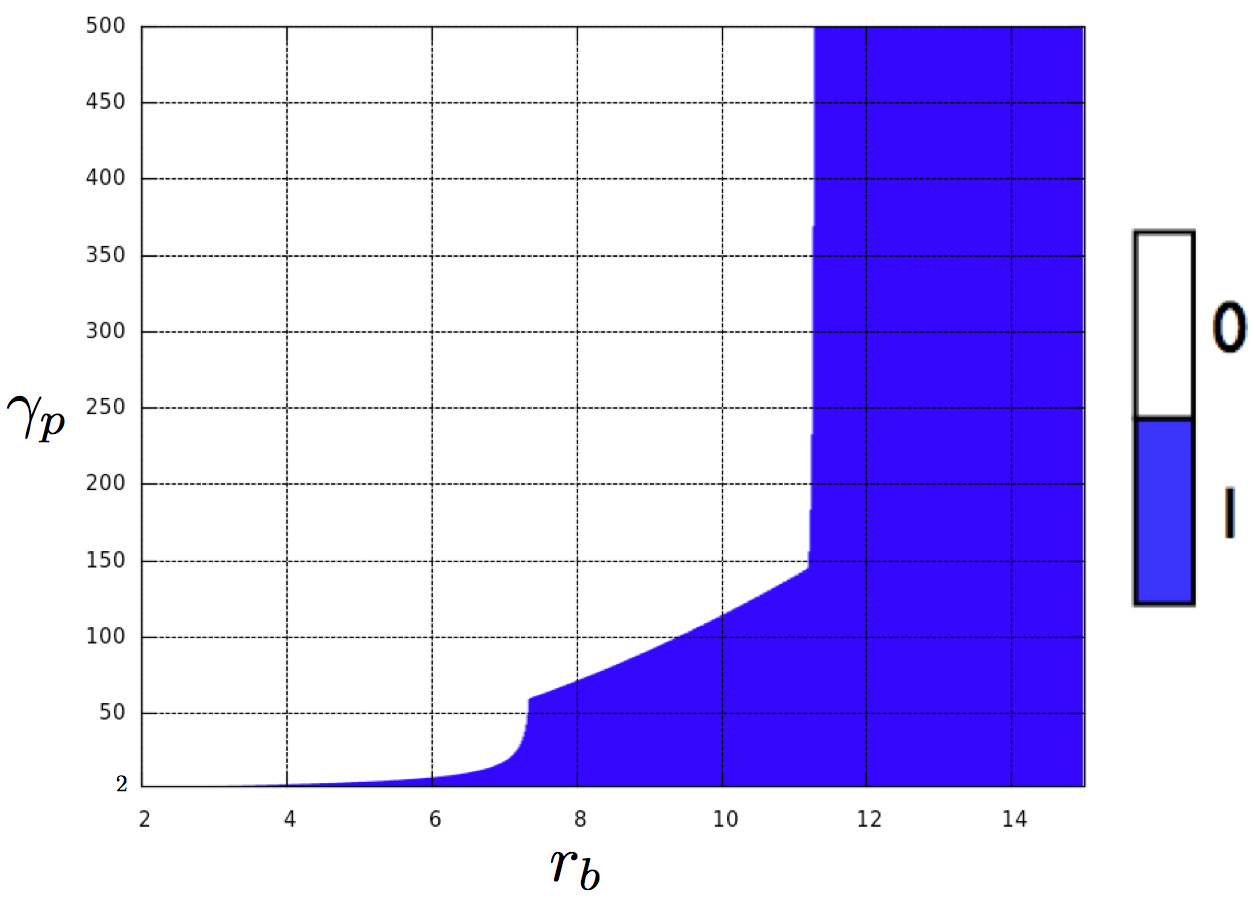}   
   \caption{For each value of $r_b$ and $\gamma_p$, a value of 1 is displayed
if $|x^{\prime}|/\gamma_p\leqslant r_b$ at the moment where $p_x^\prime=0$ for
the first time, and 0 otherwise. The initial conditions are
$x^\prime(0)=0$, $y^\prime(0)=r_b$, $\dot{x}^\prime(0)=-\gamma_pv_p$,
$\dot{y}^\prime(0)=0$ (electron at rest in the laboratory frame).}
   \label{fig5}
\end{figure}

We have performed a complete scan of the parameter space $(r_b,\gamma_p)$, as
for Fig. \ref{fig3}, but applying either the condition
$r \leqslant r_b$ or $|x^{\prime}|/\gamma_p\leqslant r_b$ at the moment where
$p_x^\prime=0$ for the first time. For the condition $r \leqslant r_b$, the
result is similar to Fig. \ref{fig3} but with a slightly different numerical
coefficient, the threshold being $r_b>0.95\gamma_p$. For the Thomas' condition,
$|x^{\prime}|/\gamma_p\leqslant r_b$, the result is non-trivial and is
displayed on Fig. \ref{fig5}.
According to this criterion, self-injection occurs for much larger values of
$\gamma_p$ than for the KNPS threshold. But there is no physical meaning for
applying $|x^{\prime}|/\gamma_p\leqslant r_b$ as a criterion for
self-injection. As we can see on Fig. \ref{fig4}, the electron streams backwards
after the point where $p_x^\prime=0$ and $|x^{\prime}|/\gamma_p\leqslant r_b$.
It should not be considered as injected in the frame of the considered model.
Therefore, the threshold for self-injection indicated in Eq. (22) of Ref.
\cite{PoP2010Thomas} is incorrect, because it relies on the elliptical
trajectory which is in contradiction with the equations used to derive the
threshold.

Nevertheless, the present discussion emphasizes that, because when $r_b$
increases $r_\textrm{max}$ becomes very close to $r_b$, a small deformation of
the bubble structure, or the consideration of the field enhancement at the back
of the bubble due to electron crossing, or the consideration of self-consistent screened fields, could considerably change the conclusion about injection or non-injection in the bubble. Considering these
effects in the model and confirming or invalidating the KNPS result are areas
for future works.

\bibliography{sebastiencorde}

\begin{thebibliography}{4}%
\makeatletter
\providecommand \@ifxundefined [1]{%
 \@ifx{#1\undefined}
}%
\providecommand \@ifnum [1]{%
 \ifnum #1\expandafter \@firstoftwo
 \else \expandafter \@secondoftwo
 \fi
}%
\providecommand \@ifx [1]{%
 \ifx #1\expandafter \@firstoftwo
 \else \expandafter \@secondoftwo
 \fi
}%
\providecommand \natexlab [1]{#1}%
\providecommand \enquote  [1]{``#1''}%
\providecommand \bibnamefont  [1]{#1}%
\providecommand \bibfnamefont [1]{#1}%
\providecommand \citenamefont [1]{#1}%
\providecommand \href@noop [0]{\@secondoftwo}%
\providecommand \href [0]{\begingroup \@sanitize@url \@href}%
\providecommand \@href[1]{\@@startlink{#1}\@@href}%
\providecommand \@@href[1]{\endgroup#1\@@endlink}%
\providecommand \@sanitize@url [0]{\catcode `\\12\catcode `\$12\catcode
  `\&12\catcode `\#12\catcode `\^12\catcode `\_12\catcode `\%12\relax}%
\providecommand \@@startlink[1]{}%
\providecommand \@@endlink[0]{}%
\providecommand \url  [0]{\begingroup\@sanitize@url \@url }%
\providecommand \@url [1]{\endgroup\@href {#1}{\urlprefix }}%
\providecommand \urlprefix  [0]{URL }%
\providecommand \Eprint [0]{\href }%
\@ifxundefined \urlstyle {%
  \providecommand \doi  [0]{\begingroup \@sanitize@url \@doi}%
  \providecommand \@doi [1]{\endgroup \@@startlink {\doibase
  #1}doi:\discretionary {}{}{}#1\@@endlink }%
}{%
  \providecommand \doi  [0]{doi:\discretionary{}{}{}\begingroup
  \urlstyle{rm}\Url }%
}%
\providecommand \doibase [0]{http://dx.doi.org/}%
\providecommand \Doi [0]{\begingroup \@sanitize@url \@Doi }%
\providecommand \@Doi  [1]{\endgroup\@@startlink{\doibase#1}\@@Doi}%
\providecommand \@@Doi [1]{#1\@@endlink}%
\providecommand \selectlanguage [0]{\@gobble}%
\providecommand \bibinfo  [0]{\@secondoftwo}%
\providecommand \bibfield  [0]{\@secondoftwo}%
\providecommand \translation [1]{[#1]}%
\providecommand \BibitemOpen [0]{}%
\providecommand \bibitemStop [0]{}%
\providecommand \bibitemNoStop [0]{.\EOS\space}%
\providecommand \EOS [0]{\spacefactor3000\relax}%
\providecommand \BibitemShut  [1]{\csname bibitem#1\endcsname}%
\bibitem [{\citenamefont {Thomas}(2010)}]{PoP2010Thomas}%
  \BibitemOpen
  \bibfield  {author} {\bibinfo {author} {\bibfnamefont {A.~G.~R.}\
  \bibnamefont {Thomas}},\ }\Doi {10.1063/1.3368678} {\bibfield  {journal}
  {\bibinfo  {journal} {Phys. Plasmas},\ }\textbf {\bibinfo {volume} {17}},\
  \bibinfo {eid} {056708} (\bibinfo {year} {2010})}\BibitemShut {NoStop}%
\bibitem [{\citenamefont {Kostyukov}\ \emph {et~al.}(2009)\citenamefont
  {Kostyukov}, \citenamefont {Nerush}, \citenamefont {Pukhov},\ and\
  \citenamefont {Seredov}}]{PRL2009Kostyukov}%
  \BibitemOpen
  \bibfield  {author} {\bibinfo {author} {\bibfnamefont {I.}~\bibnamefont
  {Kostyukov}}, \bibinfo {author} {\bibfnamefont {E.}~\bibnamefont {Nerush}},
  \bibinfo {author} {\bibfnamefont {A.}~\bibnamefont {Pukhov}}, \ and\ \bibinfo
  {author} {\bibfnamefont {V.}~\bibnamefont {Seredov}},\ }\Doi
  {10.1103/PhysRevLett.103.175003} {\bibfield  {journal} {\bibinfo  {journal}
  {Phys. Rev. Lett.},\ }\textbf {\bibinfo {volume} {103}},\ \bibinfo {pages}
  {175003} (\bibinfo {year} {2009})}\BibitemShut {NoStop}%
\bibitem [{\citenamefont {Kostyukov}\ \emph {et~al.}(2010)\citenamefont
  {Kostyukov}, \citenamefont {Nerush}, \citenamefont {Pukhov},\ and\
  \citenamefont {Seredov}}]{NJP2010Kostyukov}%
  \BibitemOpen
  \bibfield  {author} {\bibinfo {author} {\bibfnamefont {I.}~\bibnamefont
  {Kostyukov}}, \bibinfo {author} {\bibfnamefont {E.}~\bibnamefont {Nerush}},
  \bibinfo {author} {\bibfnamefont {A.}~\bibnamefont {Pukhov}}, \ and\ \bibinfo
  {author} {\bibfnamefont {V.}~\bibnamefont {Seredov}},\ }\href
  {http://stacks.iop.org/1367-2630/12/i=4/a=045009} {\bibfield  {journal}
  {\bibinfo  {journal} {New J. Phys.},\ }\textbf {\bibinfo {volume} {12}},\
  \bibinfo {pages} {045009} (\bibinfo {year} {2010})}\BibitemShut {NoStop}%
\bibitem [{\citenamefont {Kalmykov}\ \emph {et~al.}(2009)\citenamefont
  {Kalmykov}, \citenamefont {Yi}, \citenamefont {Khudik},\ and\ \citenamefont
  {Shvets}}]{PRL2009Kalmykov}%
  \BibitemOpen
  \bibfield  {author} {\bibinfo {author} {\bibfnamefont {S.}~\bibnamefont
  {Kalmykov}}, \bibinfo {author} {\bibfnamefont {S.~A.}\ \bibnamefont {Yi}},
  \bibinfo {author} {\bibfnamefont {V.}~\bibnamefont {Khudik}}, \ and\ \bibinfo
  {author} {\bibfnamefont {G.}~\bibnamefont {Shvets}},\ }\Doi
  {10.1103/PhysRevLett.103.135004} {\bibfield  {journal} {\bibinfo  {journal}
  {Phys. Rev. Lett.},\ }\textbf {\bibinfo {volume} {103}},\ \bibinfo {pages}
  {135004} (\bibinfo {year} {2009})}\BibitemShut {NoStop}%
\end{thebibliography}%

\end{document}